# Spin-transfer torques for domain walls in antiferromagnetically coupled ferrimagnets

Takaya Okuno[1], Duck-Ho Kim[1,*], Se-Hyeok Oh[2], Se Kwon Kim[3,4,*], Yuushou Hirata[1], Tomoe Nishimura[1], Woo Seung Ham[1], Yasuhiro Futakawa[5], Hiroki Yoshikawa[5], Arata Tsukamoto[5], Yaroslav Tserkovnyak[3], Yoichi Shiota[1], Takahiro Moriyama[1], Kab-Jin Kim[6], Kyung-Jin Lee[2,7,8,*], and Teruo Ono[1,9,*]

[1]Institute for Chemical Research, Kyoto University, Uji, Kyoto 611-0011, Japan

[2]Department of Nano-Semiconductor and Engineering, Korea University, Seoul 02841, Republic of Korea

[3]Department of Physics and Astronomy, University of California, Los Angeles, California 90095, USA

[4]Department of Physics and Astronomy, University of Missouri, Columbia, Missouri 65211, USA

[5]College of Science and Technology, Nihon University, Funabashi, Chiba 274-8501, Japan

[6]Department of Physics, Korea Advanced Institute of Science and Technology, Daejeon 34141, Republic of Korea

[7]Department of Materials Science & Engineering, Korea University, Seoul 02841, Republic of Korea

[8]KU-KIST Graduate School of Converging Science and Technology, Korea University, Seoul 02841, Republic of Korea

[9]Center for Spintronics Research Network (CSRN), Graduate School of Engineering Science, Osaka University, Osaka 560-8531, Japan

*E-mail: uzes.physics@gmail.com (D.-H. Kim), kimsek@missouri.edu (S. K. Kim), kj_lee@korea.ac.kr (K.-J. Lee), ono@scl.kyoto-u.ac.jp (T. Ono)

**Antiferromagnetic materials are outstanding candidates for next generation spintronic applications, because their ultrafast spin dynamics makes it possible to realize several orders of magnitude higher-speed devices than conventional ferromagnetic materials[1]. Though spin-transfer torque (STT) is a key for electrical control of spins as successfully demonstrated in ferromagnetic spintronics, experimental understanding of STT in antiferromagnets has been still lacking despite a number of pertinent theoretical studies[2-5]. Here, we report experimental results on the effects of STT on domain-wall (DW) motion in antiferromagnetically-coupled ferrimagnets. We find that non-adiabatic STT acts like a staggered magnetic field and thus can drive DWs effectively. Moreover, the non-adiabaticity parameter $\beta$ of STT is found to be significantly larger than the Gilbert damping parameter $\alpha$, challenging our conventional understanding of the non-adiabatic STT based on ferromagnets as well as leading to fast current-induced antiferromagnetic DW motion. Our study will lead to further vigorous exploration of STT for antiferromagnetic spin textures for fundamental physics on spin-charge interaction as wells for efficient electrical control of antiferromagnetic devices.**

Recently, antiferromagnets have attracted great attention because of their potential ability to serve as spintronic material platforms with features distinct from their ferromagnetic counterparts[6,7]. As neighbouring spins are aligned antiparallel in antiferromagnets, magnetic dynamics and spin transport are expected to fundamentally differ from those of ferromagnets[8]. For magnetic dynamics, recent experiments indeed found that field-driven[9] or spin-orbit-torque-driven[10,11] DW dynamics in antiferromagnetically coupled ferrimagnets is fastest at the angular momentum compensation temperature $T_A$ where the magnetic dynamics are antiferromagnetic. For spin transport, however, only theoretical studies[2-5] have investigated

STT for antiferromagnetic DWs; no experimental study has been reported yet. The theories suggest that the STT for antiferromagnetic DWs consists of adiabatic and non-adiabatic STTs, the same as for ferromagnetic DWs, and that the non-adiabatic STT acts like a staggered magnetic field; this is in contrast to the fact that the non-adiabatic STT for ferromagnetic DWs acts like a uniform magnetic field. Given that the origin of non-adiabatic STT for ferromagnetic DWs has been a subject of intensive debate[12-24] over the last decade, the nature of non-adiabatic STT for antiferromagnetic DWs is an outstanding open question. In this work, we address this question by investigating STT effects on DW motion in rare earth–transition metal (RE-TM) ferrimagnets across the $T_A$.

To obtain insight, we first describe the effect of STT on field-driven DW motion near the $T_A$ of ferrimagnets based on the collective coordinate approach. The Landau–Lifshitz–Gilbert–Slonczewski-like equation for the magnetization of a ferrimagnet is given by[19,25]

$$\delta_s \partial_t \bm{m} - \alpha s \bm{m} \times \partial_t \bm{m} - \rho \bm{m} \times \partial_t^2 \bm{m} = -\bm{m} \times \bm{h}_{\text{eff}} + P(\bm{J} \cdot \nabla)\bm{m} - \beta P \bm{m} \times (\bm{J} \cdot \nabla)\bm{m}, \quad (1)$$

where $\bm{m}$ is the unit vector along the magnetization direction, $\delta_s$ is the equilibrium net spin density along $-\bm{m}$, $\alpha > 0$ is the Gilbert damping constant, $s$ is the saturated spin density (i.e. the sum of the spin densities of two sublattices), $\rho$ is the inertia associated with the dynamics of $\bm{m}$, $\bm{h}_{\text{eff}} \equiv -\delta U/\delta \bm{m}$ is the effective field conjugate to $\bm{m}$, and $U[\bm{m}]$ is the potential energy. The last two terms on the right-hand side are the adiabatic and non-adiabatic STT terms, where $\bm{J}$ is the charge–current density, $P$ is the spin conversion factor given by $P = (\hbar/2e)(\sigma_\uparrow - \sigma_\downarrow)/(\sigma_\uparrow + \sigma_\downarrow)$ (with the electron charge $e > 0$), which represents the polarization of the spin-dependent conductivity $\sigma_s$ (s =↑ or ↓ with ↑ chosen along $-\bm{m}$) and thereby parametrizes the adiabatic torque term, and the non-adiabaticity $\beta$ characterizes the non-adiabatic torque term. The DW velocity $V$ above the Walker breakdown is obtained within the collective–coordinate approach (see Supplementary Note 1 for the derivation), given by

$$V(\mu_0 H, J) = \frac{1}{\delta_s^2 + (\alpha s)^2}(\alpha s \lambda M \mu_0 H - \delta_s PJ - \alpha s \beta PJ), \tag{2}$$

where $M$ is the magnitude of the magnetization along $\boldsymbol{m}$, $\lambda$ is the DW width, $\mu_0 H$ is the magnetic field, and $J$ is the magnitude of charge–current density. Here, the first term is the DW velocity caused by the magnetic field and the second and third terms are the contributions of adiabatic and non-adiabatic STTs to the DW velocity, respectively. When a field and a current are applied simultaneously, one can separate pure field and STT effects by adding and subtracting the DW velocity $V(\mu_0 H, +J)$ for a positive current and the DW velocity $V(\mu_0 H, -J)$ for a negative current, respectively:

$$V_{\text{field}}(\mu_0 H) = \frac{V(\mu_0 H, +J) + V(\mu_0 H, -J)}{2} = \frac{\alpha s \lambda M \mu_0 H}{\delta_s^2 + (\alpha s)^2}, \tag{3}$$

$$\Delta V(J) = \frac{V(\mu_0 H, +J) - V(\mu_0 H, -J)}{2} = -\frac{\delta_s PJ}{\delta_s^2 + (\alpha s)^2} - \frac{\alpha \beta s PJ}{\delta_s^2 + (\alpha s)^2}. \tag{4}$$

Equation (4) shows that, at the $T_A$ (i.e. $\delta_s = 0$), the adiabatic STT does not work for DW motion whereas the non-adiabatic STT acts like an effective magnetic field and thus moves the DW; this is consistent with previous theories for antiferromagnetic DWs[2-5]. We note that this effective magnetic field is *staggered* and linearly couples to the Néel order of antiferromagnetic DWs; in contrast, the non-adiabatic STT for ferromagnetic DWs is an effective *uniform* magnetic field which linearly couples to the local magnetization. Equation (4) also shows that the dependence of $\Delta V$ on $\delta_s$ is different for contributions of the adiabatic (first term) and non-adiabatic (second term) STTs. Assuming that the signs of both $P$ and $\beta$ do not change at the $T_A$ (i.e. $\delta_s = 0$), which will be justified later, the adiabatic contribution is anti-symmetric with respect to $\delta_s = 0$ whereas the non-adiabatic contribution is symmetric. This allows us to separate two STT contributions by decomposing $\Delta V$ into symmetric and anti-symmetric components.

To verify the above theoretical prediction, we investigate STT effects on current-assisted field-driven DW motion in ferrimagnetic GdFeCo compounds in which Gd and FeCo moments are coupled antiferromagnetically. Figure 1a shows a schematic illustration of our device. A 5-nm SiN/30-nm $Gd_{23.5}Fe_{66.9}Co_{9.6}$/100-nm SiN film was deposited on an intrinsic Si substrate by magnetron sputtering. The GdFeCo film was then patterned into 5-µm-wide and 500-µm-long microwires with a Hall cross structure using electron beam lithography and Ar ion milling. For current injection, 100-nm Au/5-nm Ti electrodes were stacked on the wire. As the film lacks a non-magnetic heavy metal layer as a spin-current source, the effects of spin-orbit torque[10,11,26-28] can be ignored. In this work, we focus on DW motion in the precessing regime, where the DW angle changes continuously (see Supplementary Note 2 for detailed discussion).

Figure 1b shows the DW velocity as a function of the magnetic field under positive and negative bias currents. It was found that the DW velocity linearly increases with the magnetic field and is shifted under positive and negative bias currents. The results shown in Fig. 1b can be understood in terms of Eqs. (2)–(4), i.e. $V(\mu_0 H, J) = V_{\text{field}}(\mu_0 H) + \Delta V(J)$. Following the analysis based on Eqs. (3) and (4), $V_{\text{field}}(\mu_0 H)$ and $\Delta V(J)$ can be separated.

Figure 2a shows $V_{\text{field}}(\mu_0 H = 85 \text{ mT})$ as a function of temperature $T$. It shows that $V_{\text{field}}$ reaches its maximum at 241 K, irrespective of the current density $|J| = 1.3$, 1.7, or $2.0 \times 10^{10}$ A/m², suggesting that $T_A = 241$ K [9]. Here, $T$ is a calibrated device temperature for which the Joule heating effect by the bias d.c. current is corrected (Supplementary Note 3).

Figure 2b shows $\Delta V(J)/J$ as a function of $T$. Interestingly, it was found that $\Delta V/J$ reverses its sign at 242 K, but not exactly at $T_A$ (=241 K), and is asymmetric with respect to $T$. In both figures, data points with different current densities collapse onto a single curve, justifying the Joule heating correction. Furthermore, the collapse in Fig. 2b verifies the linear dependence of $\Delta V(J)$ on $J$ in Eq. (4).

We then separate the asymmetric $\Delta V/J$ as a function of $T$ into anti-symmetric (adiabatic STT contribution) and symmetric (non-adiabatic STT contribution) components. This separation requires justification of the assumption that the signs of both $P$ and $\beta$ do not change at $T_A$. It is known that the spin polarization $p = (\sigma_\uparrow - \sigma_\downarrow)/(\sigma_\uparrow + \sigma_\downarrow)$ is determined by spin transport at the Fermi level. Therefore, $p$ is mostly determined by the FeCo sub-lattice because the 4f electron band, responsible for Gd's magnetic moment, lies much lower (about 8 eV) than the Fermi level. This assumption is further justified by an earlier experiment[29] reporting that $p$ for Co is four times larger than $p$ for Gd at a temperature below 1 K, where the Gd moment is much larger than the Co moment. Given that we use a FeCo-rich ferrimagnet, this characteristic of RE-TM ferrimagnets suggests that $T_A$ determined by integrating the net spin density over the whole Fermi sea is not special for $p$, which is determined by the spin density only near the Fermi level. Therefore, $p$, and hence the spin conversion factor $P = p\hbar/2e$, cannot change signs at $T_A$. The assumption for $\beta$ can be understood similarly. The non-adiabaticity $\beta$ originates from the spin dissipation processes, which can be caused by several distinct microscopic Hamiltonian terms independent of the net spin density; examples include spin–orbit coupling and interactions with random magnetic impurities[17,18,21]. This suggests that the sign of $\beta$ does not change at $T_A$.

In order to quantitatively compare the experimental results with the theory, we fitted the data in Fig. 2b using Eq. (4), where $\delta_s$ and $s$ were obtained by the measured $M$-$T$ curve (Supplementary Note 4) [9,30]. We considered $\alpha$, $\beta$, and $p$ as the fitting parameters, which are assumed to be temperature-independent for simplicity. The assumption of constant $\alpha$ is justified by our recent experimental report[31]. Note that this is in contrast to an earlier experiment reporting that the effective $\alpha$ of ferrimagnets, estimated from the ferromagnetic resonance (FMR) linewidth, diverges at $T_A$[32]. However, a recent theory[33] suggests that the increase of FMR linewidth at $T_A$ can be attributed to the change in the nature of the magnetic dynamics from ferromagnetic (far away from $T_A$) to antiferromagnetic (at $T_A$); not to the increase of $\alpha$ per se. The best fitting is the black solid line in Fig. 3a with the fitting parameters of $\alpha = (3.17 \pm 0.09) \times 10^{-3}$, $\beta = -0.53 \pm 0.02$, and $p = 0.109 \pm 0.002$. Figure 3b shows the adiabatic and non-adiabatic components in $\Delta V/J$ as a function of $T$, which are calculated from those parameters obtained by the fitting. The temperature dependence of the non-adiabatic component $\Delta V_{N,STT}/J$ is symmetric and that of the adiabatic component $\Delta V_{A,STT}/J$ is anti-symmetric with respect to $T_A$, as expected from Eq. (4).

Among the fitting values, a nonzero $p$ at $T_A$ is reasonable because it originates from the electron spins at the Fermi level, as explained above. On the other hand, we found a surprisingly large $|\beta|/\alpha$ of the order of 100, which is orders of magnitude larger than the values reported in most previous studies on ferromagnetic DWs[15,22,24]. To be conclusive, however, an independent test is needed for the large $|\beta|$ of antiferromagnetic DWs because $\beta$ is determined by $\Delta V/J$ at $T_A$ but, in our experiment, $T_A$ (= 241 K) is very close to the temperature ($T$ = 242 K) where the net $\Delta V/J$ is zero. We perform the following analysis as an independent estimation of $\beta$. From the field-driven current-assisted DW motion experiment, we can obtain two kinds of DW mobility: the current-induced DW mobility $\mu_C = \Delta V(J)/J$, as shown in Fig. 2b, and the field-induced DW mobility $\mu_F = V(\mu_0 H, 0)/\mu_0 H$[31]. Based on Eqs. (3) and (4), the ratio between them is expressed as

$$\frac{\mu_C}{\mu_F} = -\frac{\beta P}{\lambda M} - \frac{\delta_s P}{\alpha s \lambda M}. \quad (5)$$

The first and second terms correspond to the non-adiabatic and adiabatic STT contributions, respectively. At $T_A$ (i.e. $\delta_s = 0$), the adiabatic STT contribution disappears while the non-adiabatic contribution exists. Figure 4a shows experimentally obtained $\mu_C/\mu_F$ as a function of temperature (black). In order to clearly show the finite $\mu_C/\mu_F$ at $T_A$, we extract the symmetric (red) and the anti-symmetric (blue) components in $\mu_C/\mu_F$ as shown in Fig. 4a (Figure 4b shows the symmetric component on a magnified scale). Although the data points are somewhat scattered, using the fact that $P$ and $\beta$ should vary smoothly with $T$ near $T_A$ without any singularity, as $T_A$ of ferrimagnets is not special for these spin transport parameters, we find that the symmetric component is obviously finite in the vicinity of $T_A$. We obtain $\mu_C/\mu_F$ $(T \cong T_A)$ by averaging the data points within $T_A \pm 5$ K, which yields $3.2 \times 10^{-14}$ m²T/A; thus, $\mu_C/\mu_F$ $(T \cong T_A) = -\beta P/\lambda M = 3.2 \times 10^{-14}$ m²T/A. By substituting $M(T = T_A) = 3.9 \times 10^4$ A/m (Supplementary Note 4) and $\lambda = 15$ nm [31], we obtain $-\beta p = 0.058$. Here, the spin conversion factor $P$ is replaced by the spin polarization $p$. This value of $-\beta p$ is in good agreement with that obtained by the fitting, $-\beta p = 0.058 \pm 0.003$, supporting the quantitative validity of the fitting values.

Therefore, from both the fitting results and the mobility-based fitting-free analysis, the large $|\beta|/\alpha$ is quantitatively confirmed. As the DW velocity is proportional to $|\beta|/\alpha$, this result suggests highly efficient control of antiferromagnetic DWs by STTs. The large ratio $|\beta|/\alpha$ also has an important physical meaning. For ferromagnetic DWs, there has been longstanding debate about this ratio[12-24]. This debate is related to the underlying spin relaxation processes for $\alpha$ and $\beta$. We speculate that the observed large $|\beta|/\alpha$ in our sample originates from the spin mistracking process with the small effective exchange interaction, which has been predicted to yield a large increase in $\beta$ [16,18]. In RE-TM ferrimagnets, the effective exchange averaged over two sublattices is smaller than that of typical ferromagnetic DWs because of the antiferromagnetic alignment of RE and TM moments. We note that a peculiar feature of spin transport in RE-TM ferrimagnets has been observed recently for spin-orbit torque switching[34] but not for DW motion. Another interesting observation is the negative value of $\beta$ in GdFeCo, which is critically different from most, if not all[35], known magnets. We speculate that this negative $\beta$ may be related to the electron band structure of GdFeCo, as one theory[21] has predicted that $\beta$ can be negative in systems with both holelike and electronlike carriers.

We investigated the effects of STT on the motion of antiferromagnetic DWs in RE-TM ferrimagnets in the vicinity of $T_A$. The ferrimagnet's unique net spin density tunability via the temperature control allowed us to distinguish between the adiabatic and non-adiabatic components of STT. Our experimental results confirm the theoretical prediction of STT for antiferromagnetic DWs; the non-adiabatic torque acts like a staggered magnetic field, enabling efficient control of the DW. In addition, we found two unusual properties of STT for the antiferromagnetic DWs in the ferrimagnet. Firstly, the ratio of $|\beta|/\alpha$ is very large, which can lead to fast current-induced antiferromagnetic DW motion. Secondly, the sign of $\beta$ is negative, the origin of which demands further theoretical progress. Our work not only shows the promising utility of ferrimagnets for studying the magnetoelectronic properties of antiferromagnetically coupled systems, but also calls for further theoretical and experimental studies on STT for inhomogeneous antiferromagnetic spin textures.

## Methods

**Film preparation and device fabrication.** The studied sample was an amorphous ferrimagnetic film of 5-nm SiN/30-nm $Gd_{23.5}Fe_{66.9}Co_{9.6}$/100-nm SiN on an intrinsic Si substrate deposited by magnetron sputtering. To avoid oxidation of the GdFeCo layer, a 5-nm-thick $Si_3N_4$ layer was used as the buffer and capping layers. The film exhibits perpendicular magnetic anisotropy (PMA). GdFeCo microwires 5 µm wide and 500 µm long were fabricated using electron beam lithography and an Ar ion milling process. A negative-tone electron beam resist (maN-2403) mixed with a thinner for resists (T-1047, at a volume ratio of 1:1) was used for lithography at a fine resolution (~5 nm). For current injection, 100-nm Au/5-nm Ti electrodes were stacked on the wires.

**Experimental setup for field-driven domain wall motion.** A pulse generator (Picosecond 10,300B) was used to generate the current pulse and create the DW. A 2-mA, 100-ns current pulse was used to create the DW. For field-driven DW motion, a 2-, 2.5-, or 3-mA bias d.c. current (with a corresponding current density of 1.3, 1.7, or $2.0 \times 10^{10}$ A/m$^2$) was injected along the wire to generate an anomalous Hall voltage, $V_H$. A Yokogawa 7651 was used as a d.c. current source. The $V_H$ at the Hall cross was recorded by an oscilloscope (Tektronix 7354) through a 46-dB differential amplifier. A low-temperature probe station was used to measure the DW motion for wide ranges of temperature.

**DW detection technique.** We used time-of-flight measurement of DW propagation to obtain the DW velocity in the flow regime. The procedure for measuring the DW velocity was as follows. First, a large perpendicular magnetic field $\mu_0 H_{sat} = -150$ mT was applied to reset the magnetization. Next, a drive field $\mu_0 H_d$, in the range of $|\mu_0 H_P| < |\mu_0 H_d| < |\mu_0 H_C|$, was applied in the opposite direction. Here, $\mu_0 H_P$ is the pinning field of DW motion and $\mu_0 H_C$ is

the coercive field of the sample. Since $\mu_0 H_d$ is smaller than $\mu_0 H_C$, the drive field does not reverse the magnetization or create DWs. Next, a current pulse (2 mA, 100 ns) was injected by a pulse generator to create a DW next to the contact line through a current-induced Oersted field. We note that a 100-ns current pulse yielded a 100% DW writing probability for all temperatures and magnetic fields examined. As soon as the DW is created, the drive field $\mu_0 H_d$ pushes the DW because $\mu_0 H_d$ is larger than $\mu_0 H_P$. Then the DW propagates along the wire and passes through the Hall cross region. When the DW passes through the Hall cross, the Hall voltage changes abruptly because the magnetization state of the Hall cross reverses as a result of the DW passage. This Hall signal change is recorded by the oscilloscope through a 46-dB differential amplifier. We refer to this as a 'signal trace.' Since the detected Hall voltage change includes a large background signal, we subtract the background from the 'signal trace' by measuring a 'reference trace.' The reference trace is obtained in the same manner as the signal trace except that the saturation field direction is reversed ($\mu_0 H_{sat} = +150$ mT). In this reference trace, no DW is nucleated; hence, only electronic noise can be detected in the oscilloscope in the reference trace. The DW velocity $V$ is determined as $V = L/(t - t')$, where $L$ is the DW travel length, which was 400 μm in our measurement; $t$ is the DW arrival time; and $t'$ is the DW generation time. $t'$ was obtained as $t' = (t_{final} + t_{initial})/2$, where $t_{initial}$ ($t_{final}$) is the initial (final) time of the pulse current for DW generation. We averaged $V$ from five (when $230 < T < 250$ K) or three (otherwise) repeated measurements (note that $T_A = 241$ K in our sample). The error bar in $V$ was determined as the standard deviation of $V$ values from repeated measurements.

**Data availability.** All data that support the findings of this study are available from the corresponding authors on request.

**Figure Captions**

**Figure 1. Schematic illustration of the measurement setup and the domain wall velocity as a function of the magnetic field. a,** Schematic illustration of the device. The direction of the bias current, DW motion, and the applied magnetic field are indicated. **b,** Domain wall velocity $V$ as a function of the magnetic field $\mu_0 H$ at 211.6 K under the bias current $J = \pm\ 1.3\times10^{10}$ A/m². The dotted green line indicates $V_{\text{field}}(\mu_0 H) = [V(\mu_0 H, +J) + V(\mu_0 H, -J)]/2$ and the orange arrow indicates $2\Delta V(J) = [V(\mu_0 H, +J) - V(\mu_0 H, -J)]$.

**Figure 2. Field and current contributions to domain wall velocity as a function of temperature. a,** $V_{\text{field}}(\mu_0 H = 85\ \text{mT})$ and **b,** $\Delta V/J$ as a function of temperature under the bias current density $|J| = 1.3, 1.7,$ and $2.0\times10^{10}$ A/m².

**Figure 3. Fitting results for experimental data. a,** The experimental data $\Delta V/J$ (symbols) and the fitting result (black line). **b,** The non-adiabatic STT component $\Delta V_{\text{N,STT}}/J$ (red line) and the adiabatic STT component $\Delta V_{\text{A,STT}}/J$ (blue line) in $\Delta V/J$, which are calculated from the fitting result in Fig. 3a.

**Figure 4. Ratio between the current-induced DW mobility $\mu_C$ and the field-induced DW mobility $\mu_F$. a,** Experimentally obtained $\mu_C/\mu_F$ (black), the symmetric (red) and anti-symmetric (blue) components in $\mu_C/\mu_F$ are shown as functions of temperature. **b,** The symmetric component in $\mu_C/\mu_F$ is shown on a magnified scale. The black solid curve is provided as a guide for the eye.

**Acknowledgements**

This work was supported by the JSPS KAKENHI (Grant Numbers 15H05702, 26870300, 26870304, 26103002, 26103004, 25220604, and 2604316), Collaborative Research Program of the Institute for Chemical Research, Kyoto University, and R & D project for ICT Key Technology of MEXT from the Japan Society for the Promotion of Science (JSPS). This work was partly supported by the Cooperative Research Project Program of the Research Institute of Electrical Communication, Tohoku University. D.H.K. was supported as an Overseas Researcher under the Postdoctoral Fellowship of JSPS (Grant Number P16314). S.K.K and Y.T. acknowledge the support from the Army Research Office under Contract No. W911NF-14-1-0016. K.J.K. was supported by the National Research Foundation of Korea (NRF) grant funded by the Korea government (MSIP) (No. 2017R1C1B2009686, NRF-2016R1A5A1008184). K.J.L. acknowledges support from the National Research Foundation of Korea (NRF-2017R1A2B2006119), the Samsung Research Funding Center of Samsung Electronics under Project Number SRFCMA1702-02, and the KIST Institutional Program (Project No. 2V05750).

**Author contributions**

T.O., D.H.K., K.J.K., and T.O. planned the study. Y.F., H.Y., and A.T. grew and optimized the GdFeCo film. T.O. and D.H.K. fabricated the device and performed the experiment. Y.H., T.N., and W.S.H. helped with the experiment. S.H.O., S.K.K., Y.T., and K.J.L. provided theory. T.O., D.H.K., S.K.K., Y.S., T.M., K.J.K., K.J.L., and T.O. analysed the results. T.O., D.H.K., S.K.K., K.J.K., K.J.L., and T.O. wrote the manuscript.

**Additional information**

Correspondence and request for materials should be addressed to D.H.K., S.K.K., K.J.L., and T.O.

**Competing financial interests**

The authors declare no competing financial interests.

**a**

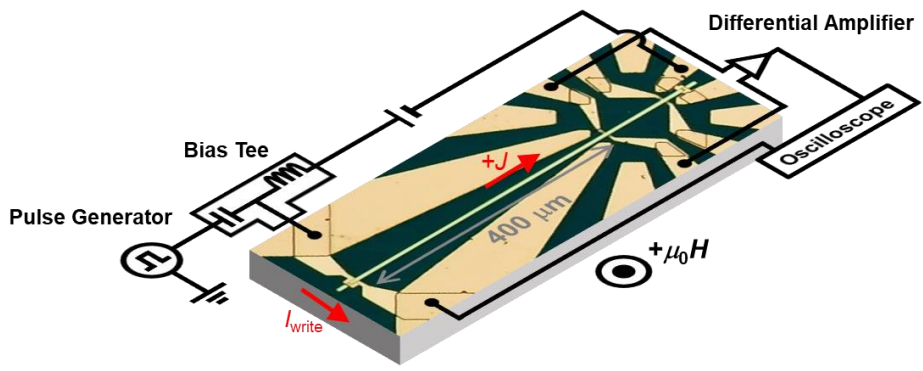

**b**

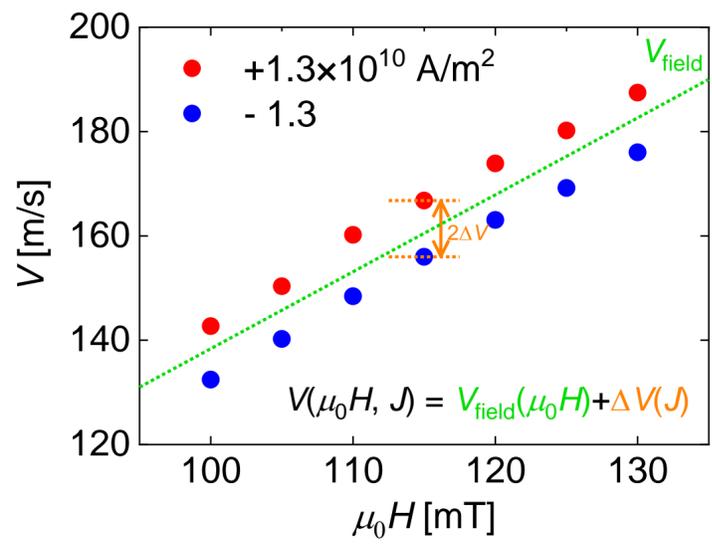

$V(\mu_0 H, J) = V_{\text{field}}(\mu_0 H) + \Delta V(J)$

**Figure 1**

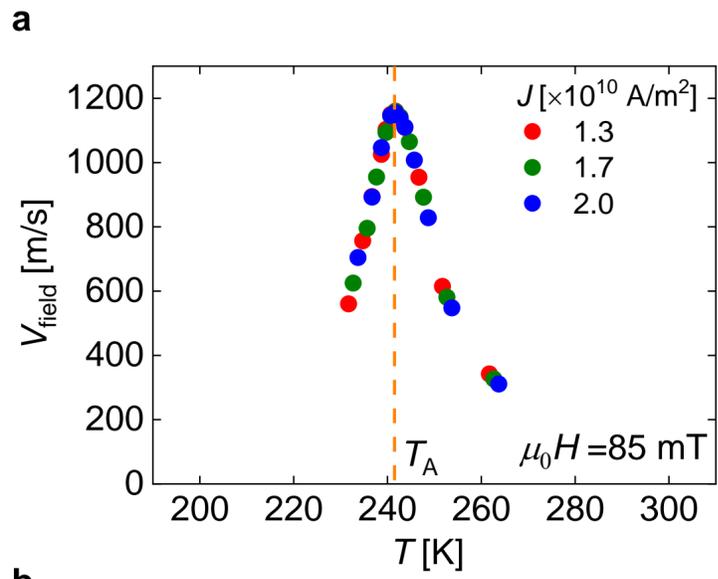

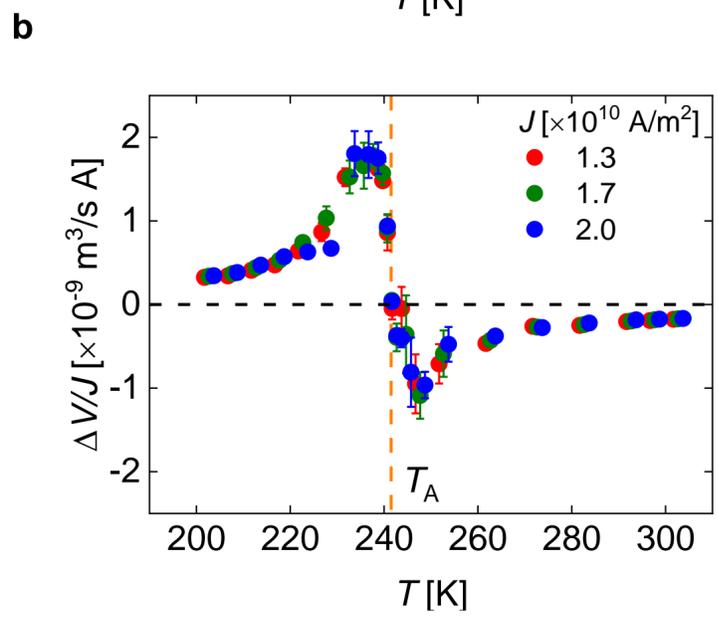

**Figure 2**

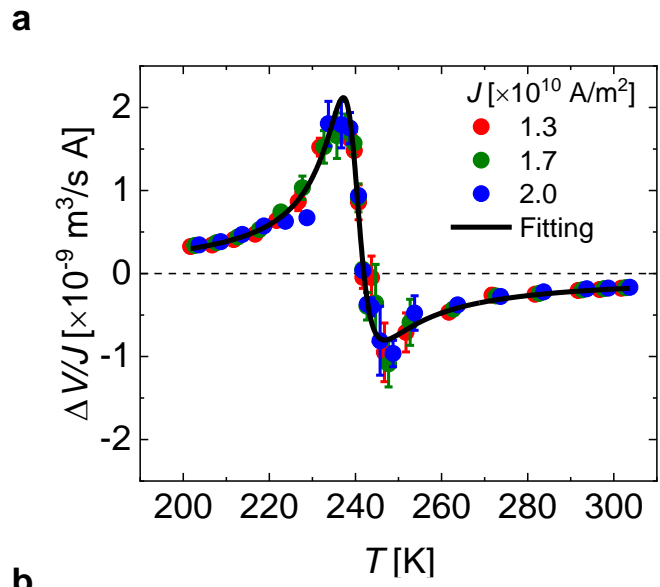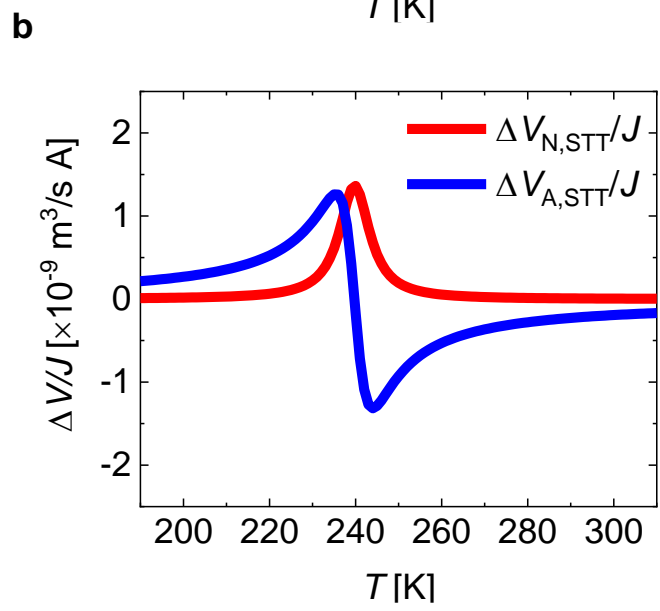

**Figure 3**

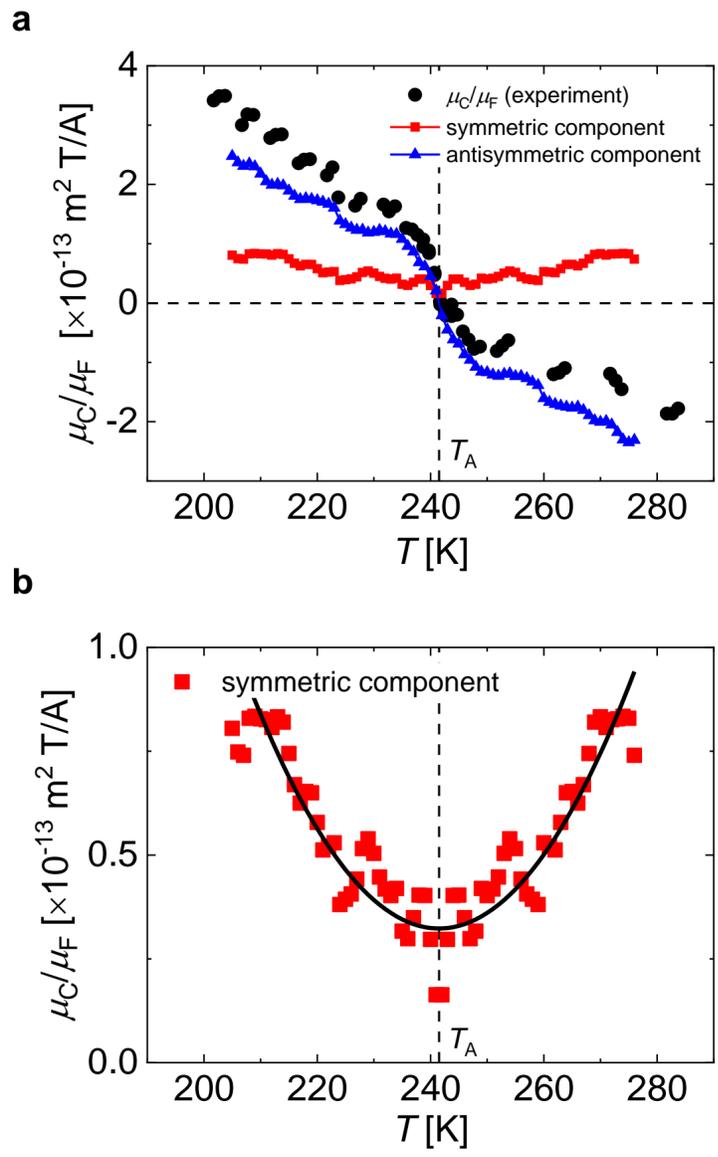

**Figure 4**